\documentclass[trackchanges]{aastex701}

\usepackage{soul} 
\usepackage{color}
\begin{document}

\title{No evidence of intrinsic linear polarization in Nova Vel 2025 (V572 Vel)}

\author[0000-0001-6013-1772]{I. J. Lima}
\affiliation{Universidade Estadual Paulista ``J\'{u}lio de Mesquita Filho'', UNESP, Campus of Guaratinguet\'{a}, Av. Dr. Ariberto Pereira da Cunha, 333 - Pedregulho, Guaratinguet\'{a} - SP, 12516-410, Brazil\\}
\email[show]{isabellima01@gmail.com}  

\author[0000-0001-7346-7638]{M. M. C. Mello}
\affiliation{Instituto Nacional de Pesquisas Espaciais (INPE/MCTI), Av. dos Astronautas, 1758, S\~ao Jos\'e dos Campos, SP, Brazil\\}
\email{marina.mello@inpe.br}

\author[0000-0002-9459-043X]{C. V. Rodrigues}
\affiliation{Instituto Nacional de Pesquisas Espaciais (INPE/MCTI), Av. dos Astronautas, 1758, S\~ao Jos\'e dos Campos, SP, Brazil\\}
\email{claudia.rodrigues@inpe.br}

\author[0000-0002-2647-4373]{G. J. M. Luna}
\affiliation{Universidad Nacional de Hurlingham (UNAHUR). Laboratorio de Investigación y Desarrollo Experimental en Computación, Av. Gdor. Vergara 2222, Villa Tesei, Buenos Aires, Argentina\\}
\affiliation{Consejo Nacional de Investigaciones Científicas y Técnicas (CONICET)\\}
\email{juan.luna@unahur.edu.ar}

\author[0000-0002-0386-2306]{F. Jablonski}
\affiliation{Instituto Nacional de Pesquisas Espaciais (INPE/MCTI), Av. dos Astronautas, 1758, S\~ao Jos\'e dos Campos, SP, Brazil\\}
\email{fjjablonski@gmail.com}

\author[0000-0002-8646-218X]{F. Falkenberg}
\affiliation{Instituto Nacional de Pesquisas Espaciais (INPE/MCTI), Av. dos Astronautas, 1758, S\~ao Jos\'e dos Campos, SP, Brazil\\}
\email{fernando.marques@inpe.br}


\begin{abstract}

We report on polarimetric observations of V572~Vel (Nova Vel 2025) conducted on June 26th and July 21st, 2025, shortly after its nova eruption was discovered. Our measurements in the I$_C$ band revealed an average linear polarization of 1.60$\pm$0.03\% at position angle of 132.2\degr. To distinguish between intrinsic and interstellar polarization, we also measured 86 field stars in the nova's vicinity, finding an average polarization of 1.25$\pm$0.61\% at a nearly identical position angle of 132\degr. The strong consistency between the nova's polarization and that of the surrounding field stars suggests that the observed polarization is predominantly interstellar in origin. We found no significant evidence of an intrinsic polarization component, which would typically arise from an asymmetric distribution of ejected material. Further multi-band observations are recommended to confirm these findings.


\end{abstract}

\keywords{ \uat{Cataclysmic variable stars}{203} --- \uat{Novae}{1127} --- \uat{Starlight polarization}{1571} --- \uat{Stars: individual (Nova Vel 2025)}{}} 


\section{Introduction} 

A nova eruption is a cataclysmic stellar event that occurs in a semi-detached binary system, where a white dwarf (WD) accretes hydrogen-rich material from a companion star, typically a main-sequence or red-giant star. When the accreted layer on the white dwarf's surface reaches a critical mass and temperature, a thermonuclear runaway is triggered, leading to a sudden and dramatic increase in the system's luminosity \citep{Shara_1989}. The geometry of the material ejected during this explosion can be anisotropic, and its shape often correlates with the nova's speed class, which is determined by how quickly its brightness declines. Slower novae tend to produce more complex, often ellipsoidal remnants, while faster novae typically yield more spherically symmetric ejecta \citep{Slavin_1995}. Light from the eruption can become linearly polarized through scattering by free electrons or dust grains within an asymmetric expanding shell \citep[e.g.,][]{Kucinskas_1990}. Consequently, linear polarimetry serves as a crucial diagnostic tool for probing the geometry of the ejecta and understanding the physical processes of the nova outburst.

V572~Vel (Nova Vel 2025) was discovered on June 25th, 2025. It is a cataclysmic variable (CV) with an orbital period of 2.956 hours \citep{2025Schaefer}. Spectroscopic follow-ups quickly confirmed its classification as a classical nova, showing prominent emission lines of H$\alpha$, H$\beta$, H$\gamma$, and Fe II \citep{2025_Shore}. X-ray observations with the Swift/XRT telescope two days post-eruption resulted in a non-detection, which is not unusual in the early, optically thick phase of a nova \citep{2025_Luna}. In this paper, we present our analysis of linear polarization observations of Nova~Vel~2025, taken one day and approximately one month after its discovery, to investigate the geometry of its ejecta.

\section{Observations and data reduction} \label{sec:obs}

The polarimetric observations were conducted on June 26th and July 21st, 2025, using the IAGPOL polarimeter \citep{Magalhaes_1996} mounted on the 0.6-m Boller and Chivens telescope at the Observatório do Pico dos Dias (OPD) operated by the Laboratório Nacional de Astrofísica (LNA), Brazil. 
The measurement process involved stepping a half-wave plate through 16 positions from 0\degr\ to 337.5\degr to modulate the incoming polarized light. The total on-source observation times were 3.7 minutes on June 26th and 38 minutes on July 21st. The polarization of field stars was measured separately on July 21st using longer exposures to ensure sufficient S/R for a large number of objects.

For calibrating purposes, observations of polarized and unpolarized standard stars were performed during the same nights. These calibrations were used to determine the instrumental reference frame for the position angle and to confirm the absence of significant instrumental polarization, which was found to be negligible. Data reduction was carried out using standard procedures in {\sc IRAF}, along with the dedicated {\sc PCCDPACK} package and custom {\sc AstroPoP} routines for polarimetric analysis.

\section{Results and conclusions} \label{sec:results}

The I$_{C}$ band magnitude of Nova Vel 2025 was measured to be 4.5 on June 26th and 5.2 on July 21st. The decline of approximately one magnitude in about one month is characteristic of a fast nova, which aligns with the expectation of more symmetric ejecta.

Our polarimetric measurements yielded a linear polarization of 1.63$\pm$0.02\% at a position angle (PA) of 131.9\degr\ on June 26th, and 1.57$\pm$0.04\% at a PA of 132.5\degr on July 21st. These values are consistent within one-sigma, demonstrating no significant variation over the one-month period. To estimate the contribution of interstellar polarization, we measured 86 field stars within a FoV of 11.3\arcmin$\times$11.3\arcmin, finding an average polarization of 1.25$\pm$0.61\% at a PA of 132\degr. The polarization of Nova Vel 2025 falls squarely within this range, strongly indicating that it is primarily interstellar in origin, caused by the dichroic extinction of light by aligned dust grains in the interstellar medium.

A Lomb-Scargle periodogram analysis of the photometric and polarimetric time-series from the longer July 21st observation revealed no significant periods which rules out rapid periodicities tied to the WD spin. Our measurements show no compelling evidence of an intrinsic polarization component; the values are stable over time and statistically indistinguishable from the surrounding field stars. This suggests that the linear polarization is dominated by the interstellar component. A negligible intrinsic polarization is consistent with the generally symmetric ejecta expected from fast novae. However, the lack of pre-eruption (quiescence) data and multi-band measurements limits a full quantitative assessment. Further observations, particularly in different photometric bands, are needed to definitively rule out a small intrinsic component and place tighter constraints on the ejecta geometry.

\begin{figure*}[ht!]
\plotone{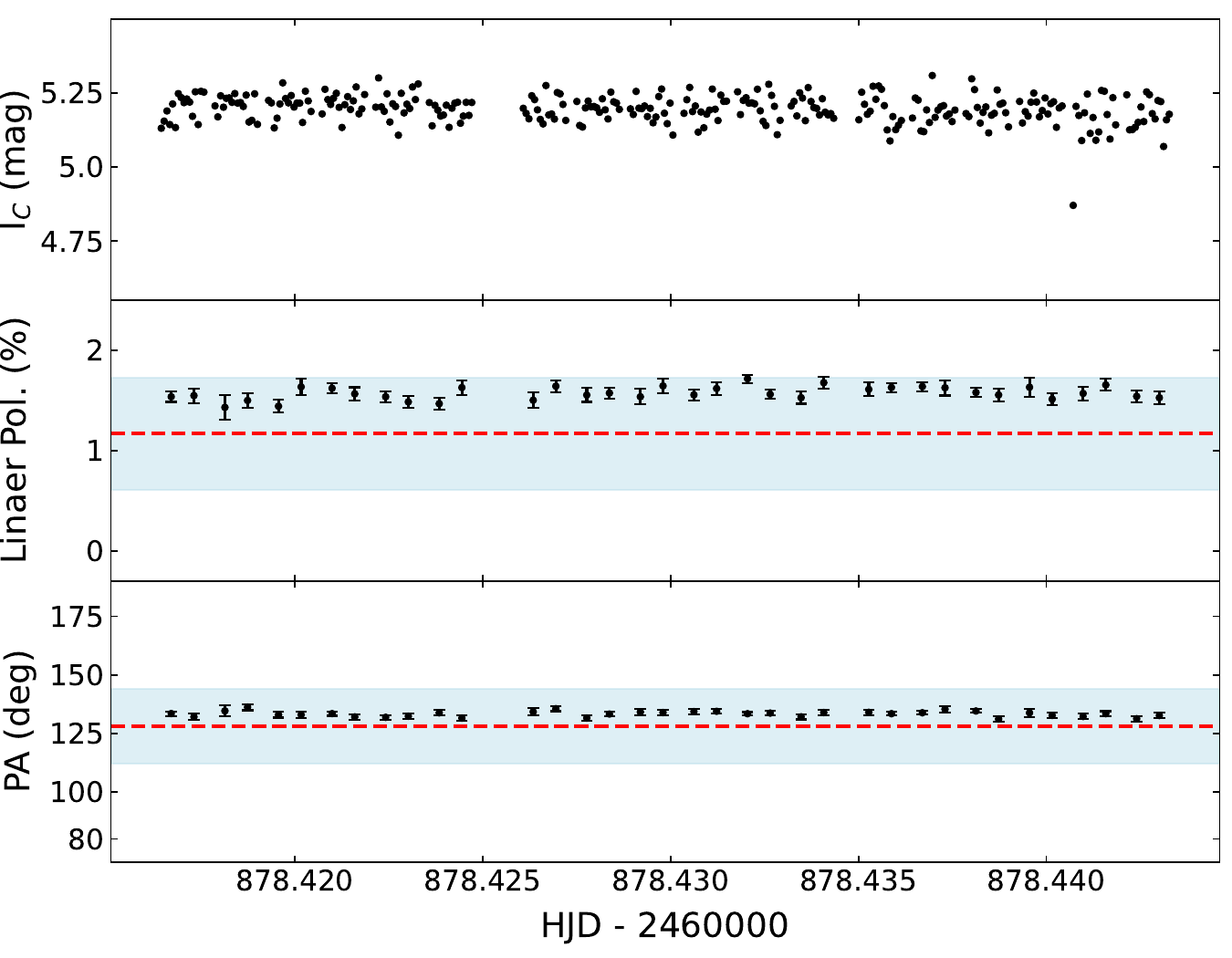}
    \caption{From top to bottom: The I$_{C}$ band magnitude, the percentage of linear polarization (\%), and the position angle (PA) of Nova Vel 2025 during the $~$38-minute continuous observation on July 21st, 2025. The solid red line in each panel indicates the average values measured for the surrounding field stars, and the light blue shaded area represents the one-sigma deviation region for those field stars.
\label{fig:1}}
\end{figure*}

\begin{acknowledgments}
We extend our gratitude to Saulo Gargaglioni and the staff at LNA for their support and for providing the necessary observation time. IJL acknowledges FAPESP for financial support through grants \#2024/14358-9 and \#2024/03736-2. CVR thanks MCTI, AEB (PO 20VB.0009), and CNPq (grant 305991/2024-8) for their funding. GJML is a member of CIC-CONICET. FF thanks CNPq (Proc: \#141350/2023-7).
\end{acknowledgments}

\begin{contribution}

All authors contributed to this manuscript.


\end{contribution}

%
\facilities{LNA:0.6m.}

\software{\mbox{IRAF} \citep{Tody/1986, Tody/1993}, \mbox{{\sc PCCDPACK\_INPE}} \citep{2018Pereyra, pereyra/2000}, \mbox{{\sc AstroPoP}} \citep{Campagnolo_2018}.
}




\bibliography{sample701}{}
\bibliographystyle{aasjournalv7}



\end{document}